\documentclass{article}    

\usepackage{graphicx}

\title{\textbf{Classical Motion}}  
\author{Richard Mould\footnote{Department of Physics and Astronomy, State University of New York, Stony Brook,
\mbox{New York} 11794-3800}}  
\date{}    
   
\begin{document}             

\maketitle              

\begin{abstract}

Preciously given rules allow conscious systems to be included in quantum mechanical systems.  There rules are derived from
the empirical experience of an observer who witnesses a quantum mechanical interaction leading to the capture of a single
particle. In the present paper it is shown that purely classical changes experienced by an observer are consistent with
these rules.  Three different interactions are considered, two of which combine classical and quantum mechanical changes. 
The previously given rules support all of these cases.    
   
\end{abstract}

\section*{Introduction}

		In a previous paper 
\cite{RM1}, \emph{drift consciousness} refers to the way the conscious attention of an observer moves to nearby
states that are distinct from the original state and from each other.  In another paper \cite{RM3}, the drift of a
\emph{conscious pulse} is one that moves continuously over brain states.  The first is characteristic of quantum mechanical
drifting, and the second is characteristic of classical drifting.  The first has been dealt with extensively in a number
of papers \cite{RM1}-\cite{RM5}, but the second needs some clarification.  

The simplest case of classical drifting (i.e., classical motion) consists of a slow but continuous change of some
conscious scene.  Imagine that an observer is looking at a red field that is slowly but continuously turning into a green
field.  At each moment of time the brain is centered on a state that has a definite color, like states $a$ and $b$ at
times $t_1$ and $t_2$ in fig.\ 1.  The shaded circle centered at point $a$ reflects the range of the \emph{conscious
pulse} at time $t_1$.  It is the observer's temporal and spatial resolution width at each moment.  Each horizontal line in
that figure represents a spatial range of states with the same color.  Neighborhood states at different times will shade
continuously from one color to another, with the leading edge of the pulse being more green, and the trailing edge being
more red.  The pulse therefore includes states that are all the same color at each moment of time, where each fades
continuously into another color as a function of time.  

\begin{figure}[t]
\centering
\includegraphics[scale=0.8]{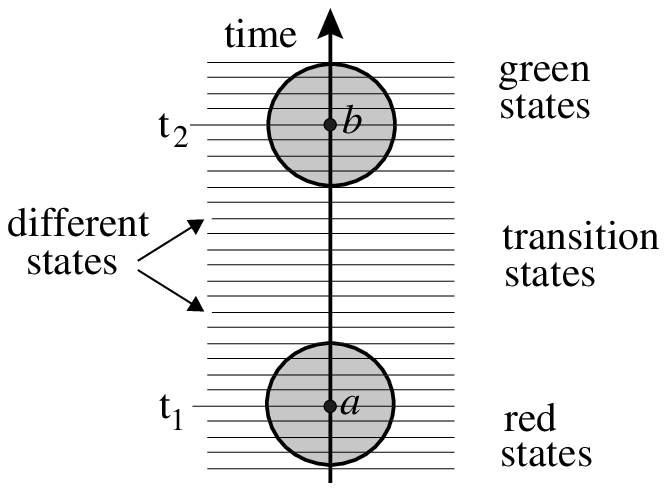}
\center{Figure 1}
\end{figure}

	If the observer's experience is more complex at each moment, then that complexity will be reflected in differences among
the horizontal states.  The most general case of classical motion is therefore represented by fig.\ 2, where each
\emph{point} in the field is a different conscious or (possibly) unconscious state.  The only requirement is that the
change from one state to the next (horizontally or vertically) is continuous.  The result will be the motion of a
conscious pulse over a classical scene that is continuous.  A classical Hamiltonian does not make discontinuous jumps even
though, in some cases, the rate of change might be too fast for the observer to resolve the states within the conscious
pulse.  

\begin{figure}[h]
\centering
\includegraphics[scale=0.8]{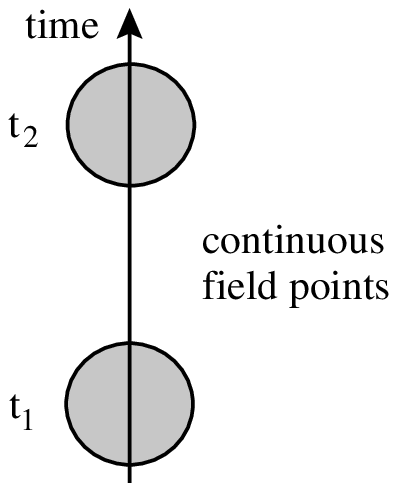}
\center{Figure 2}
\end{figure}

	One might suppose that the brain's Hamiltonian contains discontinuities that will disrupt the continuity of the field in
fig.\ 2. Brain discontinuities include abrupt changes of awareness, including periods of unconsciousness that may quickly
come or go.  However, if the brain at every moment is correlated with an environment that is classically continuous, then
the field in fig.\ 2 will be continuous.

\section*{Classical Observation of a Detector}

Prior to an observer interacting with and becoming aware of a piece of laboratory equipment such as a detector $D$, the
state of the system will take the form
\begin{displaymath}
\Phi(t_i > t) = D\{X\}
\end{displaymath}
where \{X\} is a pulse that includes unspecified conscious or unconscious brain states of the observer prior to
interaction, and $t_i$ is the time of detector/observer interaction.  When the interaction begins, the system evolves to

\begin{equation}
\Phi(t \ge t_i) = D\{X\} + .. +.. +D\{\underline{B}\}
\end{equation}
where the first component gives way to successive components that are more and more completely entangled detector and
brain states.  The succession passes continuously from the first to the last component inasmuch as there is no intrinsic
discontinuity between them.  An underlined $\underline{B}$ is a conscious brain state (refs.\ 1, 5).  The brackets around
\{X\} and \{\underline{B}\} specify a \emph{pulse} of states as defined in refs.\ 3, 5.  Ready brain states and
stochastic choices are not involved in eq.\ 1.  Rule (2) in refs.\ 1, 5 only requires the appearance of ready brain states
when there are newly emerging and ``non-continuous" brain states.  That is not the case here.  Therefore, eq.\ 1
represents a continuous drift of the observer's brain pulse that does not involve stochastic choice.  It is a purely
classical drift.

\section*{Quantum Observation}

	References 1-5 deal with discontinuities that are intrinsic to a quantum mechanical system.  The principal case studied
was that of an observer who is consciously interacting with a detector from the beginning, where the system subsequently
interacts at time $t_0$ with an incoming particle as in eq.\ 2 of \mbox{ref.\ 1} or \mbox{ref.\ 5}.  This state evolves
into two distinct components of a superposition whose detector states are \emph{not} continuous with one another.  

\begin{equation}
\Phi(t_{sc} > t \ge t_0 > t_i) = \psi(t)D_0\{\underline{B}_0\} + D_1(t)\{B_1\}
\end{equation}
where it is assumed that eq.\ 1 applies between $t_i$ and time $t_0$.  After
 $t_0$, there is continuous flow of probability current from the first component to the second in eq.\ 2.  However, unlike
the interaction in eq.\ 1, the observer continues to be conscious of the first component during this process - until there
is a stochastic hit on the second component at $t_{sc}$.  At that time the first component in this equation goes to zero
and consciousness is conferred on the ready brain \mbox{pulse $\{B_1\}$}, making it a conscious brain pulse
$\{\underline{B}_1\}$. 

This switch-over from one distinctive conscious field of states $\{\underline{B}_0\}$ to a totally different conscious
field is shown in fig.\ 3, where the initial conscious pulse is seen (on the left) to approach the interaction time $t_0$
with full strength.   After $t_0$, it weakens and an associated ready brain pulse on the right gains strength; that is,
the initial pulse loses amplitude and the ready pulse gains amplitude \footnote{This weakening does not mean that there is
a lessening of conscious intensity as one might expect from the relationship of intensity to amplitude in classical
physics.  In this case, square modulus is decreased, but the associated conscious state is undiminished.  See discussion
of this in ref.\ 3.}.  At time $t_{sc}$ the first component disappears completely and the second component becomes
conscious.

\begin{figure}[h]
\centering
\includegraphics[scale=0.8]{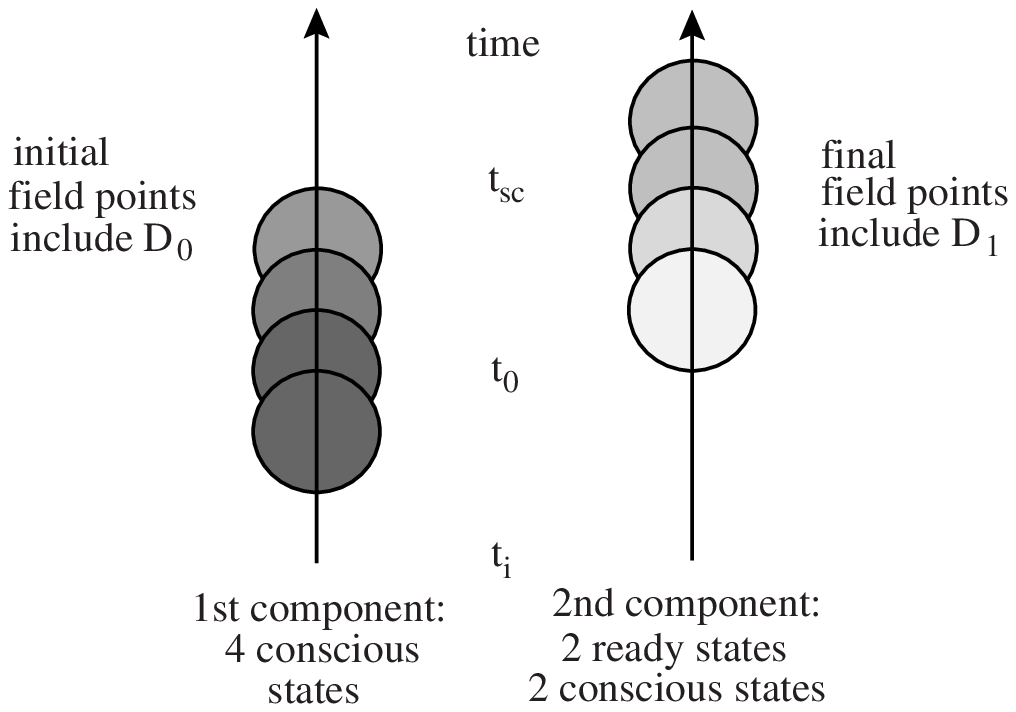}
\center{Figure 3}
\end{figure}

The instantaneous change from the first to the second component in eq.\ 2 may seem incorrect when it is noted that the
detector (being an macroscopic body) cannot undergo instantaneous change.  The detector in eq.\ 2 includes all of the low
level physiological states of the brain as well as all the process inside of the detector proper.  The time dependent
function $D_1(t)$ therefore includes all of the above changes, but the observer will only `see' the result of the changes
when the conscious pulse $\{\underline{B}_1\}$ first appears.  For the observer, all those changes will be historic
antecedents of his new experience.  That is, the \emph{final state} of the detector will be non-locally correlated with the
appearance of $\{\underline{B}_1\}$.   

To see this in detail, rewrite eq. 2 representing the detector in three parts $DDD$ where these are the early, middle, and
later parts of the detector.  Of course, the detector can be broken up into a continuum of intermediate states between
initial and final states, but we will simplify with only three.  There will then be four components in eq.\ 2. 

\begin{eqnarray}
\Phi(t_{sc} > t \ge t_0 > t_i) &=& \psi(t)D_0D_0D_0\{\underline{B_0}\} + D_1(t)D_0D_0\{B_0\}\nonumber\\
&\mbox{and}& D_1'(t)D_1D_0\{B_0\}  \hspace{.3cm}  \mbox{and}\hspace{.3cm} D_1''(t)D_1D_1\{B_1\}\nonumber
\end{eqnarray}
where the third and fourth components do not receive current because of \mbox{rule (4)}.  All of the new components are
initially zero, but the second  increases in time while the last two do not.  Since all of the current from the
first component will flow into the second component, it is certain that the stochastic trigger will select  

\begin{displaymath}
\Phi(t = t_{sc}) = D_1(t_{sc})D_0D_0\{\underline{B}_0\}
\end{displaymath}
Subsequently there will be a classical progression (given by the arrows below) that leads finally to the observer's
conscious awareness of the capture.

\begin{displaymath}
\Phi(t \ge t_{sc}) = D_1(t)D_0D_0\{\underline{B}_0\} \rightarrow D_1'(t)D_1D_0\{\underline{B}_0\} \rightarrow
D_1''(t)D_1D_1\{\underline{B}_1\}
\end{displaymath}

During the time that the capture is working its way through the detector, the observer will remain conscious of the ground
state of the detector.  He will become aware of the change to $\{\underline{B}_1\}$ only at the end.  In this and other
papers, we will skip these intermediate steps by going directly from the first to the last components as in eq.\ 2.

\section*{Pulse to Pulse}

The first component $\{\underline{B}_0\}$ in eq.\ 2 is a continuously evolving superposition of states.  The second
component $D_1(t)\{B_1\}$ is also continuously evolving, but it is a new component that is discontinuously different from
the first.  Therefore, rule (2) requires that it give rise to a \emph{ready} brain pulse $\{B_1\}$ (or perhaps a ready
pulse
$\{B_0\}$ as in the more detailed case in the previous section). The fact that each state in the conscious or ready
pulse in eq.\ 2 is embedded in a continuum of states does not negate the rule.  The Hamiltonian acts on each individual
state such as $a$ in the conscious pulse in fig.\ 4, to carry a current $J_{aa'}$ from it to the new ready state $a'$ in
that figure.   

\begin{figure}[h]
\centering
\includegraphics[scale=0.8]{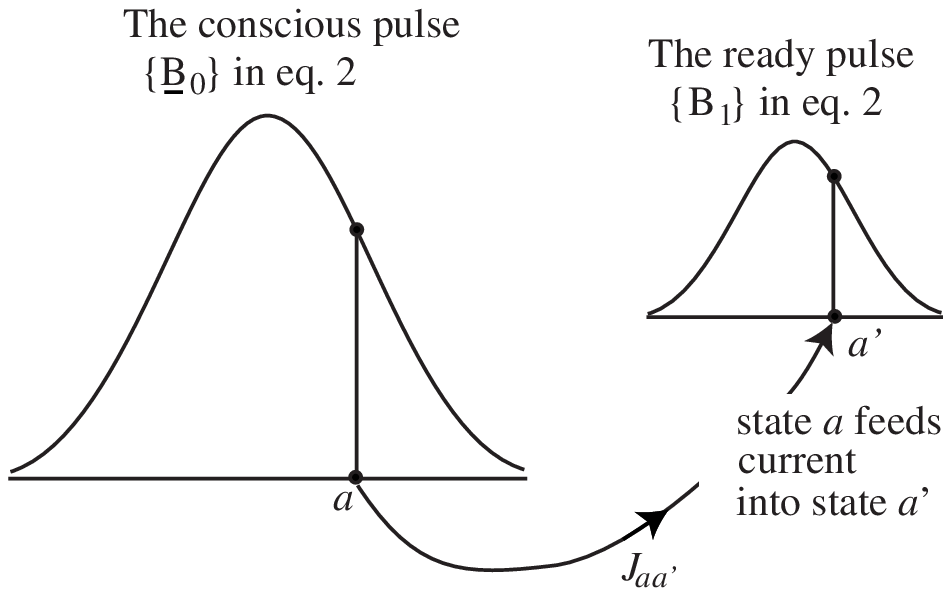}
\center{Figure 4}
\end{figure}
So state $a$ undergoes a discontinuous change to state $a'$ as required by \mbox{rule (2)}.

\section*{A Terminal Observation}

	In the case of the terminal observation described in ref.\ 1, the interaction between an incoming particle and a detector
is complete by a time $t_f$, which is assumed to be \emph{before} an observer looks at the apparatus at time $t_{ob}$.  

\begin{displaymath}
\Phi(t_{ob} > t > t_f) = [\psi(t)D_0 + D_1(t)]\{X\}
\end{displaymath}
where \{X\} is an unspecified  brain pulse of the observer that is not yet entangled with the detector.   

At time $t_i$ the interaction begins like the one in eq.\ 1, where the unspecified pulse $\{X\}$ changes
\emph{classically} into a pulse $\{\underline{B}\}$ while becoming entangled with the detector.    

\begin{displaymath}
\Phi(t \ge t_i) = [\psi(t)D_0+ D_1(t) ][\{X\} + .. + .. + \{\underline{B}\}]
\end{displaymath}
Initially, the observer will not be able to distinguish between the two superimposed detector states $D_0$ and $D_1$. 
However, at some point he will resolve the difference between them, and at this point, a continuous ``classical" evolution
will no longer be possible.  Let this happen with the first appearance of \mbox{state $\{\underline{B}\}$} at time
$t_{ob}$.  The solution will then branch `quantum mechanically' into two new components, with \mbox{rule (2)} requiring
the introduction of the ready brain pulses
$\{B_0\}$ and
$\{B_1\}$.  
\begin{eqnarray}
\Phi(t \ge t_{ob}) &=& [\psi(t)D_0 + D_1(t)]\{\underline{B}\}\\
&+& \psi'(t)D_0\{B_0\} + D_1'(t)\{B_1\}\nonumber
\end{eqnarray}
Current will  flow from $\psi(t)D_0\{\underline{B}\}$ in the first row to the first component $\psi'(t)D_0\{B_0\}$ in
the second row, and from $D_1(t)\{\underline{B}\}$ in the first row to the second component $D'_1(t)\{B_1\}$ in the
second row.  It is the flow of current to the two ready brain pulses that will lead to a stochastic choice between them. 
That choice will produce either  
 \begin{displaymath}
\Phi(t \ge t_{sc}) = \psi(t)D_0\{\underline{B}_0\} \hspace{.5cm}\mbox{or}\hspace{.5cm} \Phi(t \ge t_{sc}) =
D_1\{\underline{B}_1\}
\end{displaymath}

This evolution of the conscious field is shown in fig.\ 5.  The initial classical change $\{X\} \rightarrow
\{\underline{B}\}$ is represented by the four field points in the central column of fig.\ 5, after which
$\{\underline{B}\}$ diminishes in amplitude as current flows into the ready brain states to the right and to the left.  It
is imagined that a stochastic hit occurs on field points to the left that include the detector state $D_0$.    

\begin{figure}[h]
\centering
\includegraphics[scale=0.8]{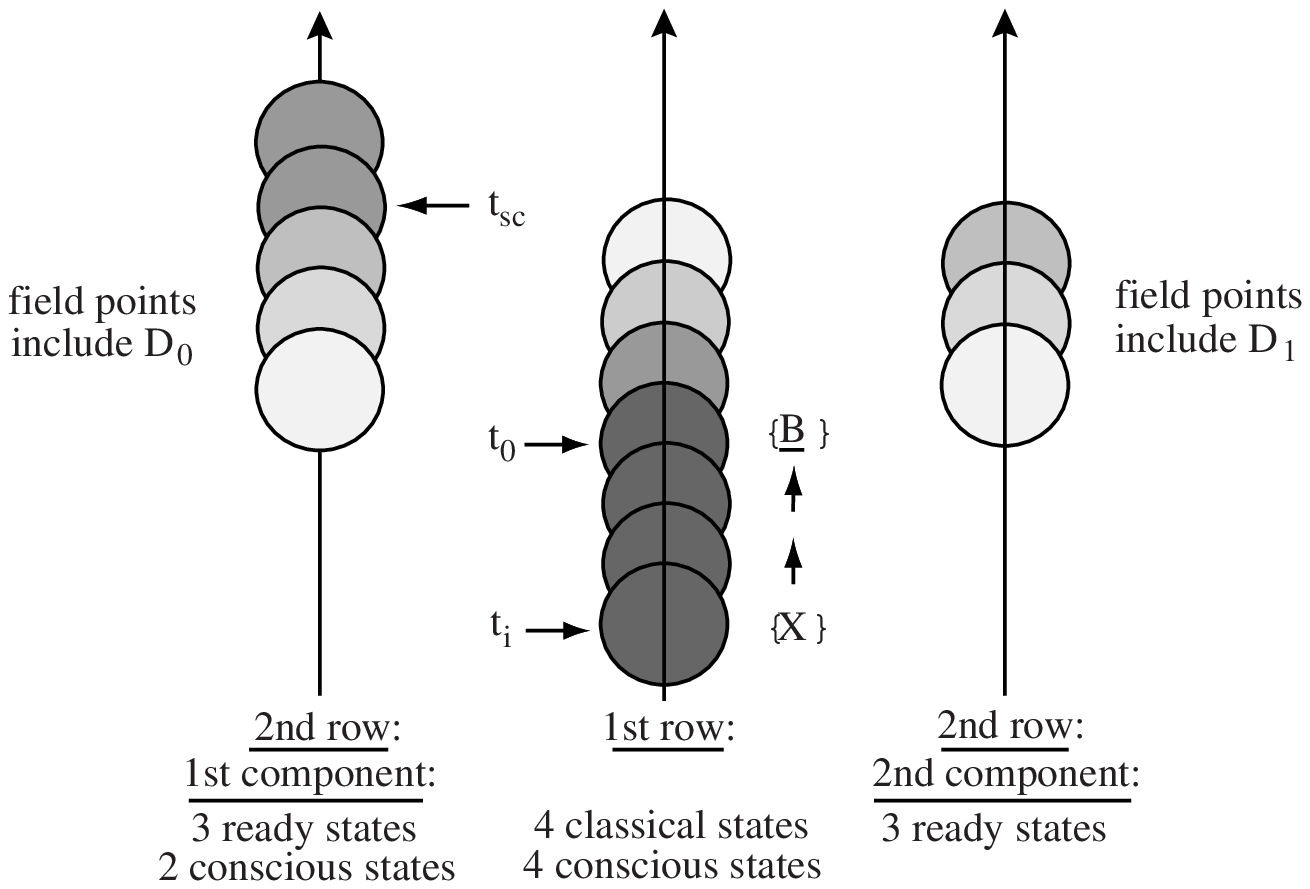}
\center{Figure 5}
\end{figure}

\section*{Conclusion}

The above examples include three very different ways that an observer can interact with a detector.  The first (eq.\ 1)
occurs when an observer interacts in a purely classical way.  The second (eq.\ 2, fig.\ 3) is a quantum mechanical
interaction that occurs after the observer is already entangled with the detector.  And the third (eq.\ 3, fig.\ 5) is a
classical/quantum mechanical interaction that occurs as the observer becomes entangled with a detector superposition.  In
the second and third cases, quantum mechanical and the classical changes are intermingled in distinctive ways.

	Generally speaking, if an observer evolves continuously along a single line in a field of possible conscious experiences,
then his motion will be classical.  A stochastic choice will not be necessary.  However, if the observer encounters a
superposition, or somehow becomes part of a superposition, then that single line will break into two or more branches. 
Since the observer cannot be conscious on more than one branch at a time, a stochastic choice must be made to decide which
branch will be followed by the conscious observer.  When a non-continuous quantum mechanical jump of this kind presents
itself, rule (2) requires that the newly emerging state must be a ready brain state.

\end{document}